\documentclass[twocolumn,amsfonts,prl,aps]{revtex4}

\usepackage{graphicx}

\begin{document}

\title{Black holes in loop quantum gravity: the complete space-time}

\author{ Rodolfo Gambini$^{1}$,
Jorge Pullin$^{2}$}
\affiliation {
1. Instituto de F\'{\i}sica, Facultad de Ciencias, 
Igu\'a 4225, esq. Mataojo, Montevideo, Uruguay. \\
2. Department of Physics and Astronomy, Louisiana State University,
Baton Rouge, LA 70803-4001}

\begin{abstract}
We consider the quantization of the complete extension of the
Schwarzschild space-time using spherically symmetric loop quantum
gravity. We find an exact solution corresponding to the semi-classical
theory. The singularity is eliminated but the space-time
still contains a horizon. Although the solution is known partially
numerically and therefore a proper global analysis is not possible,
a global structure akin to a singularity-free Reissner--Nordstr\"om
space-time including a Cauchy horizon is suggested.
\end{abstract}

\maketitle

Black holes are among the most spectacular revolutions in our
understanding of the nature of space-time that occurred as a
consequence of general relativity. In the classical theory, certain
configurations of matter cannot overcome their gravitational self
attraction and form an event horizon, a surface beyond which no
communication with the exterior is possible. Matter continues to
contract inside the horizon until a singularity is formed. Such
singularities in the theory have the desirable property of not being
able to communicate with the exterior (cosmic censorship). On the
other hand, it is expected that such singular behavior of the
classical theory could be altered significantly when one considers
quantum effects. In loop quantum gravity, for example, it is known in
the context of mini-superspace models that the big bang singularity is
eliminated and replaced by a bounce \cite{AsPaSi}. Since the interior
of a black hole is classically isometric to a Kantowski--Sachs
cosmology (that also sees its singularity eliminated in loop quantum
cosmology), it is natural to expect that the black hole singularity
will also disappear in a similar way \cite{kantowskisachs,interior}. A
complete treatment of the space-time of a black hole in loop quantum
gravity is still lacking, even within a midi-superspace type of
quantization.  The intention of this letter is to provide such a
treatment. We will consider space-times with spherical symmetry and
set up their canonical theory.  We will use a further gauge fixing to
avoid the hard problem of having structure functions in the constraint
algebra (see \cite{thiemanngiesel} for a good discussion). We will
then proceed to study classically the ``polymerized'' theory that can
be straightforwardly quantized in the loop representation. It is known
that such polymerized theories can capture many effects that one would
find in a more systematic quantization followed by a semi-classical
approximation. We will see that indeed the complete space-time can be
covered and a solution can be constructed that replaces the
singularities (black and white hole) of the usual Kruskal diagram by
regular surfaces. We will show that in fact such surfaces can be
smoothly matched so where one expected a ``black hole'' one tunnels
into a ``white hole'' region of another universe and this can be
continued indefinitely. The resulting solution therefore has a Cauchy
horizon and can be characterized as the analog in semiclassical loop
quantum gravity of an eternal black hole.

We will use the Ashtekar new variables to describe the spherically
symmetric space-times. Previous work on this subject was done in
modern language by Bojowald and Swiderski \cite{boswi} so we refer the
reader to them for details.  There is only one non-trivial spatial
direction (the radial) which we call $x$ since it is not necessarily
parameterized by the usual radial coordinate. We will elaborate more
on the range of $x$ later.  The canonical variables usual in loop
quantum gravity are a set of triads $E^a_i$ and $SO(3)$ connections
$A_a^i$; after the imposition of spherical symmetry one is left with
three pairs of canonical variables
$(\eta,P^\eta,A_\varphi,E^\varphi,A_x,E^x)$.  The variables $\theta$
and $\varphi$ are angles transverse to the radial direction as in
usual polar coordinates. Instead of using triads in the transverse
directions, one introduces a ``polar'' set of variables
$E^\varphi,\eta$ and their canonical momenta. It is convenient to
introduce the gauge invariant variable $K_x$ defined by $2\gamma K_x =
A_x +\eta'$ and also $K_\varphi$ defined as $A_\varphi=2\gamma
K_\varphi$, where $\gamma$ is the Immirzi parameter of loop quantum
gravity. . The canonically conjugate pairs are now $E^x,K_x$ and
$E^\varphi,K_\varphi$. The relationship to more traditional metric
variables is,
\begin{eqnarray}
g_{xx}&=& \frac{(E^\varphi)^2}{|E^x|},\qquad g_{\theta\theta} = |E^x|,\\
K_{xx}&=&- K_x {\rm sign}(E^x) \frac{(E^\varphi)^2}{\sqrt{|E^x|}},
\quad
K_{\theta\theta} = -\sqrt{|E^x|} K_\varphi,\nonumber
\end{eqnarray}
and the latter two are the components of the extrinsic curvature.  The
diffeomorphism and Hamiltonian constraints can be seen in detail in
ref. \cite{exterior}.  These constraints have the usual constraint
algebra for gravity in $1+1$ dimensions, which includes structure
functions. This implies the usual ``problem of dynamics'' of canonical
quantum gravity.  Our strategy to treat this model will be to bring it
down to a model with one Abelian constraint and a true Hamiltonian.
That way it can be treated using the standard Dirac procedure. To
achieve this we eliminate the diffeomorphism constraint by choosing a
gauge that determines the functional form for $E^x=f(x,t)$. Imposing the
constraint strongly determines $K_x$. This also fixes the
corresponding Lagrange multiplier (the shift)
$N^r=-\dot{f}(x,t)/f'(x,t)$ and also breaks reparametrization
invariance. One is left with a theory with a single constraint that is
Abelian and with a true Hamiltonian, the dynamical variables are
$E^\varphi$ and $K_\varphi$ and the constraint is,
constraint,
\begin{equation}
\Phi = -\sqrt{E^x}-K^2_\varphi \sqrt{E^x}+\frac{1}{4} 
\frac{\left(\left(E^x\right)'\right)^2\sqrt{E^x}}{\left(E^\varphi
\right)^2}+2 M \label{constraint}
\end{equation}
with $M$ an integration constant and the evolution is given by a true
Hamiltonian,
\begin{equation}
H_{\rm true}= 
 \int dx \frac{\dot{f}(x,t)}{f'(x,t)}
E^\varphi \left(K_\varphi\right)',
\end{equation}
which preserves the constraint upon evolution. We assume the spatial
manifold (the radial direction) has two boundaries. The theory at the
boundary can be constructed in similar fashion as in the exterior 
case so we refer the reader to \cite{exterior} for reasons of space.
One ends up with one degree of freedom in the boundary (the mass) that
does not evolve in time and coincides with the constant $M$.

The quantization of the Abelian constraint is straightforward and can
be carried out in the same Hilbert space we considered in the exterior
case. In brief, one discretizes the radial direction and the Hilbert
space is a tensor products of a Hilbert space of loop quantum
cosmology per spatial point. In such a space the constraint
(\ref{constraint}) is not well defined, but one can work with an
expression where $K_\varphi$ is replaced by $\sin(\mu K_\varphi)/\mu$.
The latter is immediately expressible in terms of holonomies and
therefore naturally exists in the loop representation. The resulting
theory agrees with general relativity in the limit $\mu\to 0$. In loop
quantum gravity it is natural to consider a finite value of $\mu$,
usually associated with the elementary quantum of area \cite{AsPaSi}.

Instead of quantizing the theory and then studying the semiclassical
limit, we will follow a procedure that is known \cite{interior} to
capture some of the semiclassical behaviors, in particular the
elimination of the singularity, at least in simple examples with
constant value of $\mu$ as the one we are considering.   We
analyze the resulting classical ``polymerized'' theory with finite $\mu$.
One then considers a classical theory of gravitation, different from
general relativity that contains some of the ingredients of the
quantum theory, akin to when one works out in an effective theory.

We wish to choose the function $f(t,x)$ in such a way that in the
limit $\mu\to 0$ one recovers the standard Schwarzschild metric
in Kruskal-like coordinates. That is, a metric with a singularity at
$x^2-t^2=-1$. On the other hand, in the case of finite $\mu$ we
would like the surface $x^2-t^2=-1$ to correspond to a regular
surface of the metric beyond which the metric can be extended.
To be more specific we will choose $E^x=f(u,t,\delta)$, where
$u=x^2-t^2+1$ and $\delta(\mu)$ a positive parameter such that
when  $\mu\to 0$, $\delta\to 0$ and we recover the standard
Kruskal form of the Schwarzschild space-time. To completely
fix the gauge and obtain an explicit solution 
we set $K_\varphi=g(u,t,\delta)$ after polymerization. 
In the quantum theory such a 
gauge fixing would be equivalent to the study of an evolving
constant \cite{rovelli,gapo} 
$E^\varphi$ in terms of c-number variable $K_\varphi$.

We will require the following conditions on the gauge fixing. We
choose the $u$ in the range $[0,\infty]$ and is such that the radial
variable has a logarithmic dependence on $u$, $r=\sqrt{E^x}\sim
M\ln(u)$ for $u\to \infty$. Moreover, asymptotically $E^\varphi\sim
r+M$ in ordinary Schwarzschild coordinates, which appropriately
transformed is $E^\varphi \sim (2M/\sqrt(u))(M\ln(u)+M)$.  The conjugate
variables are exponentially small in the radial coordinate 
$K_x\sim K_\varphi\sim
1/\sqrt{u}$. These boundary conditions are very similar to those in
Kruskal coordinates \cite{varadarajan}.  We did not choose to work
exactly in Kruskal coordinates asymptotically given the complicated
relation between $r$ and $u$ in those coordinates.  At $u=0$ we will
require that all variables be $t$-independent and we will choose their
derivatives to vanish (in the case of $K_\varphi$ we choose the
derivative of $\sin(\mu K_\varphi)$ to vanish, since it is the
relevant expression for the determination of the metric components via
the constraints.  This ensures that one can easily continue the
manifold without shells of matter present at $u=0$. There might be
other possibilities for this boundary condition but we have not
explored them. Finally we would like that in the limit $\delta\to 0$
we get a gauge choice that covers the entire extension of the
Schwarzschild space-time (as we mentioned, it will not be exactly the
same as the Kruskal extension, but related to it via non-singular, yet
complicated, coordinate transformations).  Although the choice of
coordinates we are making is not unique, it is computationally
laborious to actually find a coordinate system that satisfies all the
conditions we listed and that involves variables that do not turn
complex in certain regions and that has the variable $K_\varphi$
taking correct values in the Bohr compactification.

The specific choice we make for $E^x$ is,
\begin{eqnarray}
E^x&=&\left\{
\frac
{\left[\delta\left(1+u\right)+\left(10 u^2+u^{7/2}\right)
\left(\delta\left(t^2-1\right)+1\right)\right]}
{u^{7/2}+\left(t^2-1\right)\left(\delta u^{7/2}+\delta^2\right)
+\frac{1}{2}\delta^2 u
}\right.\nonumber\\
&&\left.\times\left[
\ln(1+u)\right]^2+\delta^8\right\}M^2.
\end{eqnarray}
This choice has the property that for $u\to
0$ $E^x=M^2\delta^8$ independent of $t$, for large values of $u$ it
behaves as $M^2\left(\ln^2(u)+\delta^8\right)+O(u^{-1})$, 
in the limit $\delta\to0$ we have
that $E^x= M^2 (10 u^{3/2}+u^3)\ln^2(1+u)/u^3$ tends to $0$ when
$u=0$, as in the Kruskal coordinates, giving rise to the singularity.
It can be checked that the first derivative with respect to $x$ of
$E^x$ vanishes for $u=0$ for any finite value of $\delta$. This choice
for $E^x$ is not unique, in the sense that other choices may satisfy
the above conditions. It might be possible to find simpler choices.

For $K_\varphi$ we choose,
\begin{eqnarray}
&&K_\varphi= 
\frac{1}{2}\frac{\delta^{5/2}\pi\left(1+\ln\left(1+u^2\right)\right)}
{\mu\left(\delta^{5/2}+\ln\left(1+u\right)^2\right)}
+\frac{|t|\ln\left(1+u^3\right)}{u^{3/8}}\nonumber\\
&&\times
\frac{
\left(-1+\frac{u}{\left(10+u\ln\left(1+u\right)\right)}
+\frac{\left(1+8\right)u}{\left(100+u\ln\left(1+u\right)^2\right)}\right)}
{\left(\delta^2 t+\ln\left(1+u^3\right)\right)
\left(1+u^{1/8}\right)}.
\end{eqnarray}

This choice has the property that for $u\to 0$ $K_\varphi=\pi/(2\mu)$
independent of $t$, so the term that appears in the Hamiltonian goes
as $\sin\left(\mu K_\varphi\right)\sim 1$.  This means that the
departure of the polymerized theory from continuum general relativity
is maximum at the point where the singularity would have occurred in
the continuum theory. Therefore loop quantum gravity could remove the
classical Schwarzschild singularity.  In the limit $\delta\to0$ we
have that $K_\varphi$ blows up when $u=0$, as in the Kruskal
coordinates, also compatible with the presence of the singularity in
the continuum theory.  For large values of $u$ $K_\varphi$ behaves as
$t/\sqrt{u}+O(u^{-1/8})$, It can be checked that the first derivative
with respect to $x$ of $\sin(\mu K_\varphi)$ vanishes for $u=0$. As in
the case of $E^x$, the choice is not unique.  It should also be noted
that the choice is only valid in $|t|>1$. We have extended the
solution beyond that domain. The extension is symmetric under $t\to-t,
x\to-x$, but it makes the expressions too lengthy, so for reasons of
space here we concentrate in the region $|t|> 1$ since it includes the
singularity.

\begin{figure}[ht]
\includegraphics[height=6cm]{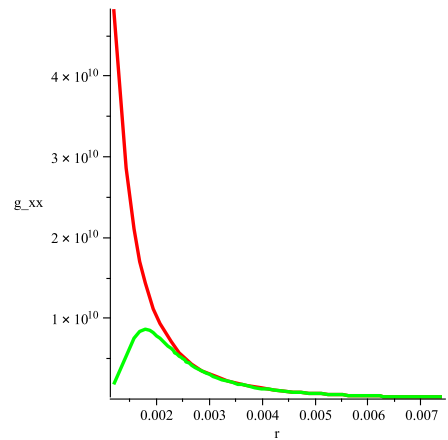}
 \caption{The metric component $g_{xx}$ shown as a function of $r$ for
 the usual Schwarzschild solution and the solution of the polymerized
 theory. The plots are for $\delta=10^{-42} $ nd $M=1$.  One sees the
 two graphs coincide long before one reaches the horizon at $r=2M$,
 but inside the Schwarzschild solution tends to blow up at $r=0$
 whereas the solution of the polymerized theory becomes finite. Close
 to the origin the plotting program cannot capture that the solution
 of the polymerized theory grows again and takes a large, but finite,
 value.  Although the comparison with Schwarzschild is suggestive, one
 must exercise care since we are plotting a coordinate dependent
 quantity in two different theories. One can show that both curves
 will agree in the limit $\delta\to 0$. The behavior of components of
 the curvature tensor grows monotonically as $r\to 0$ up to a maximum
 a finite value at the tunneling in the polymerized theory.}
\label{gxx}
\end{figure}

We would now like to generate a solution to the constraint and the
evolution equations of the polymerized theory.  We will adopt the
following strategy: we solve the diffeomorphism constraint for $K_x$
and the remaining constraint for $E^\varphi$.  The preservation of the
gauge conditions in time determine the lapse and shift. The
consistency of the system, that is, the preservation of the
constraints upon the Hamiltonian evolution guarantees that the
evolution equations for the canonical variables are automatically
satisfied.

We start by obtaining $E^\varphi$ from the Hamiltonian constraint,
which is immediate since the relation is algebraic,
\begin{equation}\label{evarphi}
E^\varphi = \frac{1}{2} \left(E^x\right)'\left(
{\sqrt{1 -\frac{2M}{\sqrt{E^x}} 
+\frac{\sin\left(\mu K_\varphi\right)^2}{\mu^2}}}
\right)^{-1},
\end{equation}
recall that $(E^x)'$ is given by (4). In these expressions prime
means derivative with respect to $x$.

Since $\left(E^x\right)'$ vanishes for $u=0$ and one wishes 
$E^\varphi$ to be finite there to avoid having a singularity, one
needs the denominator of (\ref{evarphi}) to vanish. This 
condition determines the relation between $\mu$ and $\delta$,
$\mu =\frac{\delta^2}{\sqrt{{2}- \delta^4}}$.

It is worthwhile showing explicitly the behavior of $E^\varphi$ as
$u\to 0$, $ E^\varphi|_{u\to0} = 
2M^2\delta^{11/16}+(\delta^{1/8}\left(120(t^2-1)-1\right)+120)M^2u^2
\delta^{9/16}$
which confirms that
$\left(E^\varphi\right)'=0$ at $u=0$, one of the conditions we wanted.
The behavior at large $u$ for the metric is given by,
\begin{equation}
g_{xx}|_{u\to\infty}=4 \frac{M^2 }{u}+8\frac{M^2 }{u\ln(u)},
\end{equation}
and this is just the coordinate transform of $(1+2M/r)$ with $r=M  \ln(u)$ 
to leading orders in asymptotic powers of $u$, yielding the familiar form
of the Schwarzschild solution to leading order in $1/r$.

We now proceed to determine the lapse, using the conservation in time of 
the second gauge condition, the one involving $K_\varphi$. We compute
$\dot{K}_\varphi$ using the total Hamiltonian. From there one immediately 
gets
\begin{equation}
N'= -\frac{1}{4} \frac{\dot{K}_\varphi \left(E^x\right)'-K'_\varphi \dot{E}^x}
{\left(1-\frac{2M}{\sqrt{E^x}} +\frac{\sin\left(\mu K_\varphi\right)^2}{\mu^2}
\right)^{3/2}}
\end{equation}
and via a quadrature one obtains $N$. We were not able to compute the
expression for the latter in closed form, but numerical evaluations
are straightforward. It is good to get asymptotic expressions for the
integrals to use as boundary data for the numerical integrals.  We can
therefore reconstruct all components of the space-time metric. We can
use this to study the causal structure of light cones. With this we
can locate the horizon by studying at each value of $t$ the radial
position at which the hypersurface tangent to $\sqrt{E^x}={\rm
const.}$ becomes null. We have carried out the numerical computations
for values $t=[10,100]$. For larger values the computation becomes
harder due to numerical issues.  We will choose the parameter
$\delta=10^{-8}$ and $10^{-42}$ to study convergence. To understand
the meaning of these values it is worthwhile noticing that the ratio
between the radius at the point where the curvature takes its maximum
value and the Schwarzschild radius to be of the order of
$\delta^{1/14}$ Since we are dealing with the classical polymerized
theory there is no notion of Planck mass. Using estimates based on the
treatment of the interior using the Kantowski--Sachs isometry and that
the polymerized theory departs from general relativity in scales
associated with the Planck length, one can conclude that one would be
dealing with a black hole of $3-1000$ Planck masses for both choices
of $\delta$ we make. Corrections with respect to the usual
Schwarzschild solution at the position of the event horizon are of the
order $\delta^{1/2}$, i.e. for the choices we make from $10^{-4}$ to
too small to be detected with the accuracy we are working.  
\begin{figure}[ht]
\includegraphics[height=4cm]{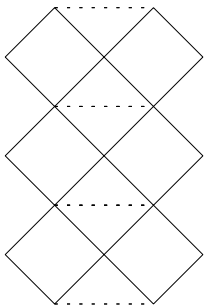}
 \caption{The conjectured global structure of the solution.
   The singularity is replaced by a regular region indicated with a
   dashed line. The space-time is continued through into another copy
   of the same solution. The solution would have a Cauchy horizon similar
   to that in a Reissner--Nordst\"om solution, presumably unstable.}
\end{figure}

Summarizing, we have carried out a midi-superspace treatment of
spherically symmetric space-times in loop quantum gravity. We have
studied a classical solution that captures the features of the
semiclassical theory. The singularity is avoided and a picture is
suggested in which the spacetime of a (highly idealized) eternal black
hole is continued into another region containing a Cauchy horizon,
similar to a Reissner--Nordstr\"om space-time but without the
singularity. In spite of the lack of singularity, there still is a
horizon and a causal behavior far away from the singularity similar to
that of the usual Schwarzschild solution. 

Is the solution unique? At this point we cannot say. There clearly are
parameters that can be changed, but it is not clear if they just
correspond to diffeomorphisms. Although the treatment of the exterior 
carried out previously
\cite{exterior}
yields a single solution up to diffeomorphisms, it is known that in
the treatments of the interior the ``polymerization'' breaks
Birkhoff's theorem
\cite{kantowskisachs,interior} 
suggesting it may not hold in the complete case
either. In the interior treatment there appears an additional
parameter in the solution which, for instance, controls if the
``bounce'' is symmetric or not and the extent of the region where the
polymerized theory departs from general relativity. Our solution
appears to have several free parameters, even though we have imposed
by hand that the bounce be symmetric.  Clarifying the uniqueness point
may shed light on the degrees of freedom that are remnant of the
elimination of the singularity in loop quantum gravity and may yield a
picture with elements in common with the ``fuzzballs'' \cite{fuzz} of
string theory, although our solutions do not exhibit significant
departures from general relativity at the position of the horizon.

We wish to thank Abhay Ashtekar and Rafael Porto for comments
and Miguel Campiglia for help with early calculations.
This work was supported in part by grants NSF-PHY0650715,
and by funds of the Horace
C. Hearne Jr. Institute for Theoretical Physics, FQXi, PEDECIBA
 and PDT \#63/076 (Uruguay) and CCT-LSU.

\end{document}